\documentclass[aps,prl,preprint,showpacs]{revtex4}
\usepackage{dcolumn}
\usepackage{bm}
\usepackage{latexsym}
\usepackage{color,graphicx}
\begin{document}

\thispagestyle{empty}

{\baselineskip0pt
\leftline{\baselineskip14pt\sl\vbox to0pt{
              \hbox{\it Yukawa Institute for Theoretical Physics}
             \hbox{\it Kyoto University}
              \vspace{1mm}
             \hbox{\it Department of Mathematics and Physics}
              \hbox{\it Osaka City  University}
              \vss}}
\rightline{\baselineskip16pt\rm\vbox to20pt{
           {
           \hbox{YITP-11-20}
           \hbox{OCU-PHYS-347}
           \hbox{AP-GR-90}
           }
\vss}}%
}

\vskip1cm

\title{Is super-Planckian physics visible?\\
{\it -- Scattering of black holes in 5 dimensions--}}
\author{
Hirotada Okawa$^{1}$
\footnote{Electronic address: okawa@yukawa.kyoto-u.ac.jp},
Ken-ichi Nakao$^{2}$
\footnote{Electronic address: knakao@sci.osaka-cu.ac.jp}
and
Masaru Shibata$^{1}$
}
\affiliation{
$^{1}$Yukawa Institute for Theoretical Physics, Kyoto University,
Kyoto 606-8502, Japan. \\
$^{2}$Department of Mathematics and Physics,
Graduate School of Science, Osaka City University,
Osaka 558-8585, Japan.
}
\date{\today}

\begin{abstract}                
It may be widely believed that probing short-distance physics is
limited by the presence of the Planck energy scale above which scale
any information is cloaked behind a horizon.  If this hypothesis is
correct, we could observe quantum behavior of gravity only through a
black hole of Planck mass.
We numerically show that in a scattering of two black holes in the
5-dimensional spacetime, a visible domain, whose curvature radius is
much shorter than the Planck length, can be formed. Our result
indicates that super-Planckian phenomena may be observed without an
obstruction by horizon formation in particle accelerators.

\end{abstract}


\maketitle
\section{Introduction}

It is well known that if quantum effects are taken into account, a
threshold energy scale, over which general relativity looses its
validity, emerges. This threshold is called the Planck scale.
For the 4-dimensional spacetime, the Planck energy is defined by
$E_{\rm pl}:=\sqrt{\hbar c^5 /2G}=1.1\times 10^{19}$~GeV, where $c$,
$\hbar$, and $G$ are the speed of light, Dirac constant, and Newton's
gravitational constant, respectively.  The circumferential radius of a
spherically symmetric black hole of mass $E_{\rm pl}c^{-2}$ is equal
to its reduced Compton wavelength $\hbar cE_{\rm pl}^{-1}$, and hence
such a black hole will behave as a gravitating quantum object whose
behavior is unpredictable in the framework of general relativity.  The
Planck energy, $E_{\rm pl}$, is much larger than the electroweak scale
($\simeq 100$~GeV), and this fact is recognized as the hierarchy
problem in the elementary particle physics.

The large extra-dimension scenario was proposed as a solution for the
hierarchy problem~\cite{Arkani-hamed,RS}. This scenario is inspired by
superstring theories, and in this scenario, the fundamental Planck
energy, $E_P$, may be as low as $10^3$~GeV scale.  The peculiarity of
this scenario is that the length scales of the compactification of
extra-dimensions can be much larger than the fundamental Planck length
$\hbar c E^{-1}_P$.  Hence, the gravity can be described by a 
higher-dimensional classical theory (e.g., higher-dimensional general
relativity) for the distance scale smaller than the compactification
scale and larger than $\hbar c E_P^{-1}$. Also in this scenario, the
classical theory of gravity will loose its validity in the
super-Planckian domains.

For non-gravitational interactions, a length scale (e.g., Compton
wavelength), which is usually explored by a particle scattering,
decreases with increasing its energy measured at the center of mass
frame. However, this relation may not hold for the phenomena in which
gravity plays an important role.  It is widely believed that 
collisions of particles with the super-Planckian energy scale would 
produce black holes, and hence, physical processes characterized by 
the length scale shorter than $\hbar cE_P^{-1}$ are hidden inside 
black holes~\cite{GT01}.  This is a kind of the cosmic censorship
hypothesis for the super-Planckian domain, but it is not trivial
whether this hypothesis is correct.  The original cosmic censorship
hypothesis claims that, roughly speaking, naked singularities are not
formed in our universe~\cite{penrose1969}. However, we have to note
that even if the cosmic censorship hypothesis is correct, it does not
necessarily imply that the super-Planckian physics is hidden inside
horizons.  In the framework of general relativity or in the large
extra-dimension scenario, the super-Planckian curvature is not a
spacetime singularity as long as it is finite, and thus the presence
of the super-Planckian domains may be irrelevant to the cosmic
censorship. Indeed, Nakao, Harada, and Miyamoto (NHM) recently
suggested, by a simple dimensional analysis, a possibility that
visible super-Planckian domains may be produced in high-energy
particle collisions, if the spacetime dimension is larger than
four~\cite{NHM2010}.

To theoretically explore the phenomena in a higher-dimensional gravity,
numerical relativity is probably the unique approach.  In the past a
few years, several implementations for the higher-dimensional
numerical relativity have been
developed~\cite{SY09,Zil,LP,Sorkin,Witek}, and now, it is feasible to
explore the nonlinear physics such as high-velocity collision in
higher dimensions as in 4 dimensions~\cite{SCPBG,shibata_OY,SCPBHY}.
In this paper, we numerically show that in a scattering of two black
holes in the framework of 5-dimensional (5D) general relativity,
super-Planckian domains may be indeed visible.

Hereafter, we adopt the natural units $c=\hbar=1$ and the
abstract index notation: Latin indices except for $w,~x,~y,~z$
denote a tensor type and Greek indices denote a component
with respect to some basis vectors~\cite{Wald}.
In this paper, we define the fundamental Planck energy $E_P$
such that 5D Einstein's equation is written as
\begin{equation}
G_{ab}=3\pi^2 E_P^{-3}T_{ab}. \label{einstein}
\end{equation}

\section{Super-Planckian Domain}

We define a super-Planckian domain as a region where 5D general
relativity looses its validity.  In the case that quantum
effects on a spacetime geometry are not very large, quantum corrections
to the Einstein-Hilbert action might be given in the form of scalar
polynomials of the Riemann tensor. Hence, we adopt the square root of
the absolute value of the Kretschmann invariant
$|R^{abcd}R_{abcd}|^{1/2}$ as a reference quantity.

To determine a reasonable threshold value of
$|R^{abcd}R_{abcd}|^{1/2}$ over which the domain becomes
super-Planckian, we adopt, as a reference, the 5D
Schwarzschild-Tangherlini (ST) black hole which is a spherically
symmetric vacuum solution in 5D general relativity.  By the definition
of $E_P$ through Eq.~(\ref{einstein}), the circumferential radius of
the ST black hole with mass $M$ is equal to $(M/E_P^3)^{1\over2}$.
Thus, the ST black hole with $M=E_P$ will behave as a gravitating
quantum object which actually cannot be described by general
relativity, because the reduced Compton wavelength of this black hole
agrees with its circumferential radius.  $|R^{abcd}R_{abcd}|^{1/2}$ at
the event horizon of this black hole is equal to $6\sqrt{2} E_P^2$.
Hence, by introducing a dimensionless reference quantity ${\cal
K}\equiv (6\sqrt{2}E_P^2)^{-1}|R^{abcd}R_{abcd}|^{1/2}$, we can define
a super-Planckian domain $\cal A$ by
\begin{equation}
\inf_{\cal A}{\cal K} \geq 1. \label{criterion}
\end{equation}
Note that the above condition will be one of sufficient
conditions for the appearance of a super-Planckian domain, because a
domain, in which one of scalar quantities defined from the Riemann
tensor exceeds an appropriately determined critical value, should be
regarded as a super-Planckian one~\cite{NHM2010}.

\section{Initial data}

Hereafter, we consider a scattering of two non-rotating black
holes with identical mass $M$.  A procedure for setting initial data
of the scattering problem with negligible junk radiation was presented
in~\cite{shibata_OY}, which we follow.

When the distance between two black holes is much larger than their
gravitational radii $R_{\rm g}:=(E_P^{-3}M)^{1/2}$, the metric near
each black hole in each rest frame is well approximated by that of the
ST black-hole spacetime,
\begin{equation}
ds^2=-\alpha^2(r_0)dt_0^2+\psi^2(r_0)\left(dw_0^2+dx_0^2+dy_0^2+dz_0^2\right),
\end{equation}
where
\begin{equation}
\psi(r)=1+\left(\frac{R_{\rm g}}{2r}\right)^2
~~~~{\rm and}~~~~
\alpha(r)=\frac{2-\psi(r)}{\psi(r)},
\end{equation}
and $r_0=\sqrt{w_0^2+x_0^2+y_0^2+z_0^2}$. To obtain an approximate
metric in the vicinity of each black hole in the {\em center-of-mass
frame} of the two-black-hole system, we perform coordinate
transformations for the above metric twice. First, a boost
transformation, $t=\Gamma(t_0\mp vw_0)$, $w=\Gamma(w_0\mp vt_0)$,
$x=x_0$, $y=y_0$, and $z=z_0$, is performed, where the velocity $v$ is
a positive constant less than unity, and $\Gamma=1/\sqrt{1-v^2}$ is
the Lorentz factor. Next, a spatial translation, $w\rightarrow w
\mp \ell_w$ and $x\rightarrow x \mp \ell_x$, is performed, where 
$\ell_w$ and $\ell_x$ are positive constants, respectively.  As a
result, we obtain two coordinate systems in which the world line of
the black hole (puncture) is given by $w=\pm(\ell_w-vt)$,
$x=\pm\ell_x$, $y=z=0$, and the line element is
\begin{eqnarray}
&&ds_\pm^2=-\Gamma^2\left(\alpha_\pm^2-v^2\psi_\pm^2\right)dt^2
\pm 2\Gamma^2v\left(\alpha_\pm^2-\psi_\pm^2\right)dtdw \nonumber \\
&&~~~~~~~~~~+\psi_\pm^2\left(B_\pm^2dw^2+dx^2+dy^2+dz^2\right),
\end{eqnarray}
where $\alpha_\pm = \alpha(r_\pm)$, $\psi_\pm =\psi(r_\pm)$, 
and $B_\pm^2 :=\Gamma^2\left(1-v^2\alpha_\pm^2\psi_\pm^{-2}\right)$
with $r_\pm:= \sqrt{\Gamma^2(w\mp\ell_w\pm vt)^2+(x\mp\ell_x)^2+y^2+z^2}$.
The components of the extrinsic curvature of the spacelike hypersurface 
labeled by $t$ are
\begin{eqnarray}
K^\pm_{ww}&=&\mp\frac{v\Gamma^3(w\mp\ell_w\pm vt)B_\pm}{r_\pm} \nonumber \\
&&~~~~~~\left[2\alpha_\pm' -\frac{\alpha_\pm}{2} 
\left\{\ln(\psi_\pm^2-v^2\alpha_\pm^2)\right\}'\right],\\
K^\pm_{xx}&=&K^{\pm}_{yy}=K^{\pm}_{zz}
=\mp\frac{\Gamma v(w\mp\ell_w\pm vt)\alpha_\pm\psi_\pm'}{B_0\psi_\pm r_\pm}, \\
K^\pm_{wx}&=&\mp\frac{\Gamma v(x\mp\ell_x)B_\pm}{r_\pm}\left[\alpha_\pm'
-\frac{\alpha_\pm}{2}\left\{\ln(\psi_\pm^2-v^2\alpha_\pm^2)\right\}'\right], \\
K^\pm_{wy}&=&\mp\frac{\Gamma vyB_\pm}{r_\pm}\left[\alpha_\pm'
-\frac{\alpha_\pm}{2}\left\{\ln(\psi_\pm^2-v^2\alpha_\pm^2)\right\}'\right], \\
K^\pm_{wz}&=&\mp\frac{\Gamma vzB_\pm}{r_\pm}\left[\alpha_\pm'
-\frac{\alpha_\pm}{2}\left\{\ln(\psi_\pm^2-v^2\alpha_\pm^2)\right\}'\right],
\end{eqnarray}
and the other components vanish. Here, the prime denotes the
ordinary derivative with respect to $r_\pm$.  Based on the above
results, we write the initial data for a scattering of two black
holes with initial velocities $\pm v$.  The metric of the
spacelike hypersurface is written in the following form
\begin{equation}
dl^2=(\Psi+\Phi)^2(B^2dw^2+dx^2+dy^2+dz^2),
\label{4-metric}
\end{equation}
where
\begin{eqnarray}
&&\Psi=1
+\left(\frac{R_{\rm g}}{2r_+}\right)^2
+\left(\frac{R_{\rm g}}{2r_-}\right)^2, \\
&&B^2=\Gamma^2\left[1-\frac{v^2}{\Psi^{4}}(1-2\Psi)^2\right].
\end{eqnarray}
The extrinsic curvature is written by 
\begin{equation}
K_{ab}=K^{+}_{ab}+K^-_{ab}+\delta K_{ab}.\label{eq13}
\end{equation}
Finally, we set $t=0$.  The unknown functions $\Phi$ in
Eq.~(\ref{4-metric}) and $\delta K_{ab}$ in Eq.~(\ref{eq13}) should be
determined from the conditions that the Hamiltonian and momentum
constraints are satisfied.  However, if the coordinate separation
between two black holes, $2\sqrt{\ell_w^2+\ell_x^2}$, is much larger
than $R_{\rm g}$, $|\Phi|$ and $|\delta K_{\mu\nu}|$ are much smaller
than $|\Psi|$ and $|K^{\pm}_{\mu\nu}|$~\cite{shibata_OY}. We choose
the initial separation to be sufficiently large and set the small
corrections to be zero. Such approximation is acceptable for our
present purpose.

\section{Numerical Results}

Numerical simulations were performed using {\small SACRA-ND} code
reported in Ref.~\cite{SY09}, in which the so-called
Baumgarte-Shapiro-Shibata-Nakamura formalism~\cite{BSSN} together with
the ``moving puncture'' approach~\cite{puncture1} and adaptive mesh
refinement algorithm~\cite{sacra} are employed.  The numerical
accuracy is monitored by computing $L2$-norm of the Hamiltonian and
momentum constraints.  The convergence of the numerical solution was
tested varying the grid resolutions with the grid spacing as
$\Delta/R_{\rm g}=15/320$, 12/320, and 10/320. We
confirmed a reasonable convergence behavior (see below).

\begin{figure}[t]
 \includegraphics[scale=0.75]{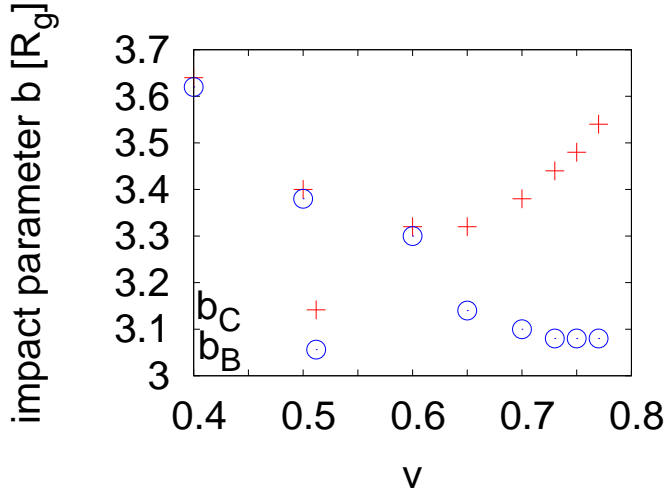} \label{fig:fig1}
 \caption{Impact parameters $b_{\rm B}$ and $b_{\rm C}$ are plotted as
 functions of the initial speed $v$ of each black hole. For $v< 0.6$,
 $b_{\rm B}$ and $b_{\rm C}$ agree with the critical impact parameter
 $b_{\rm crit}$. }
\end{figure}

Numerical simulations were systematically performed varying $v$ and an
impact parameter defined by $b\equiv2\ell_x$.  For a small velocity $v
\alt 0.6$, we were always able to determine a critical value of the
impact parameter, $b=b_{\rm crit}$, for the merger of two black
holes. Namely, for $b < b_{\rm crit}$, two black holes merge to a
single spinning black hole whereas for $b > b_{\rm crit}$, two black
holes go apart to infinity after one scattering.  The zoom-whirl orbit
was never found in 5 dimensions in contrast to the 4D
case~\cite{shibata_OY}, because the law of the gravitational force is
modified: Note that in the Newtonian limit in 5 dimensions, both
gravitational and centrifugal forces are proportional to $r^{-3}$
where $r$ is the separation of two objects.

For a high velocity with $v > 0.6$, by contrast, we were not able to
determine the value of $b_{\rm crit}$ using our current code. The
reason is that for $b_{\rm B} < b < b_{\rm C}$, the numerical
simulation crashed soon after the scattering of two black holes
occurs.  Figure 1 plots $b_{\rm B}$ and $b_{\rm C}$ as functions of
$v$ (note that for $v \leq 0.6$, $b_{\rm B}=b_{\rm C}=b_{\rm
crit}$). However, we were able to confirm that for $b < b_{\rm B}$,
two black holes merges, while for $b > b_{\rm C}$, the merger does not
happen and two black holes merely go apart to infinity after the
scattering.  In this paper, we focus on the case $b \geq b_{\rm C}$.

\begin{figure*}[t]
\begin{tabular}{ccc}
\begin{minipage}{0.33\textwidth}
 \includegraphics[scale=0.53]{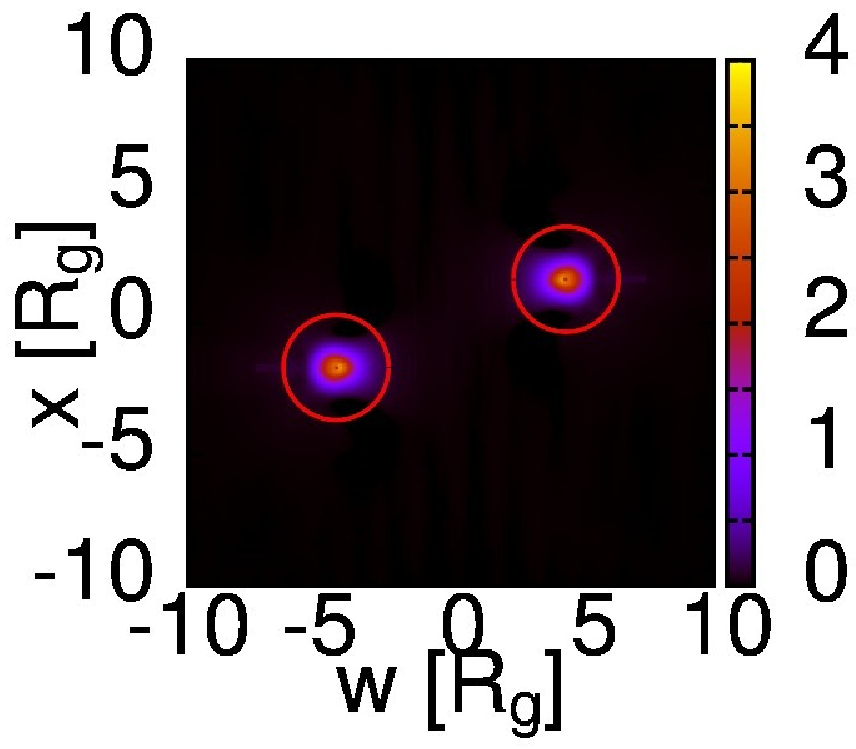}
\end{minipage}
 &
\begin{minipage}{0.36\textwidth}
 \includegraphics[scale=0.55]{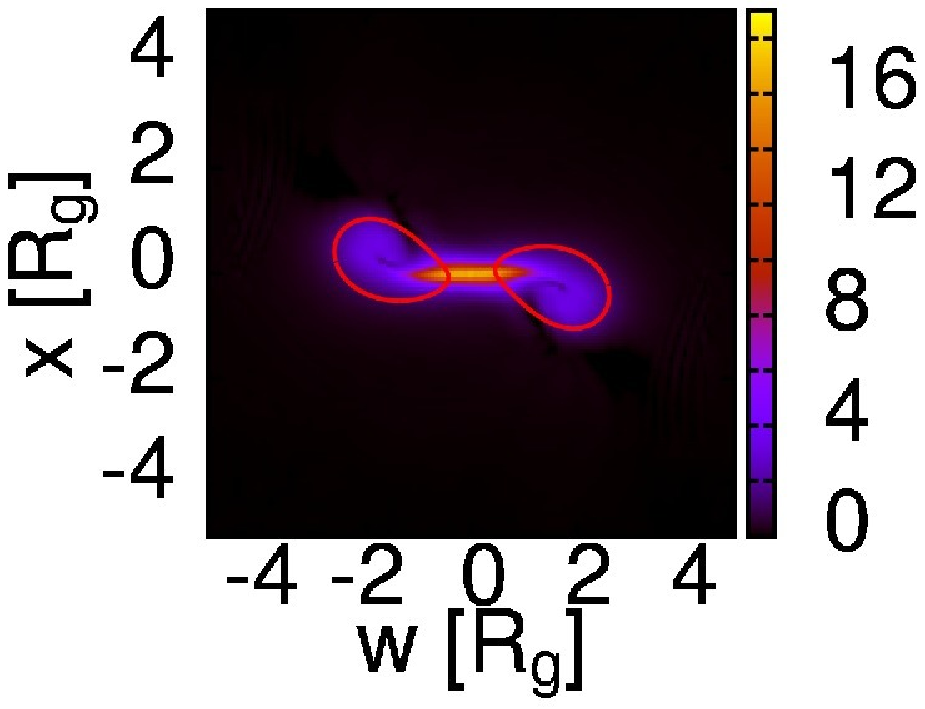}
\end{minipage}
 &
\begin{minipage}{0.33\textwidth}
 \includegraphics[scale=0.64]{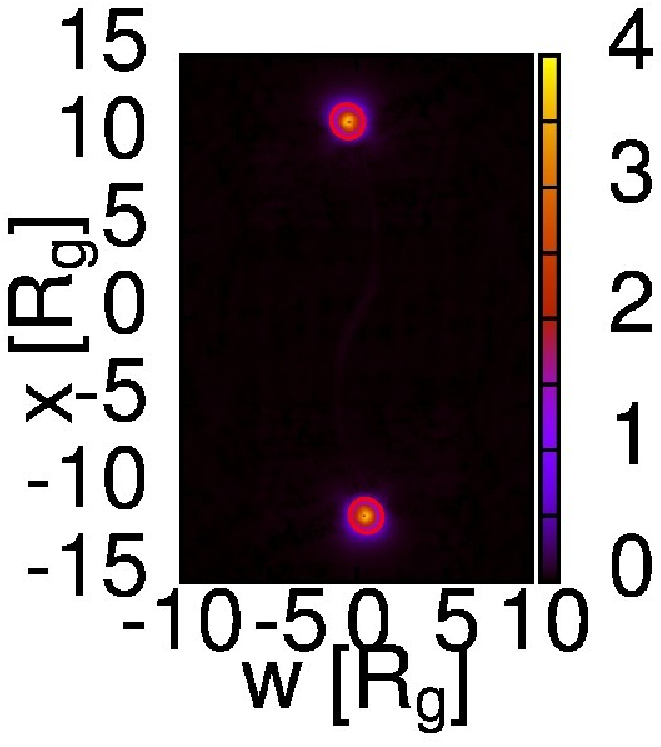}
\end{minipage}\\
 (a)&
 (b)&
 (c)
\end{tabular}
 \label{fig:fig2} \caption{Color maps of ${\cal K}$ in the scattering
 of two black holes with $v=0.7$ and $b=3.38R_g$ ; (a) before the
 scattering, (b) during the scattering, and (c) after the scattering.
 At the stage (b), a highly elongated domain with a large value of
 $\cal K \gg E_P/M$ appears between two black holes.  The solid distorted
 circles denote the apparent horizon of the black holes.  }
\end{figure*}

Figure~2 plots the time variation of $\cal K$ in a scattering process
with $v=0.7$ and $b=3.38R_{\rm g}$. The value of $\cal K$ is shown in
the unit of $E_P M^{-1}$ in the $w$-$x$ plane. In this figure, the
apparent horizon of each black hole is denoted by the solid circles.
When the separation between two black holes is much larger than
$R_{\rm g}$, no super-Planckian domain emerges outside the apparent
horizons as long as the mass of each black hole $M$ is larger than
$E_P$ [see Fig.~2(a)].  By contrast, when the separation is equal to
2--$3R_{\rm g}$, a highly elongated domain with a large value of $\cal
K \gg E_P/M$ is formed between two black holes [see Fig.~2(b)].
Finally, two black holes are scattered away, and the domain with a
large value of $\cal K$ disappears [see Fig.~2(c)].

\begin{figure}[t]
 \includegraphics[scale=0.7]{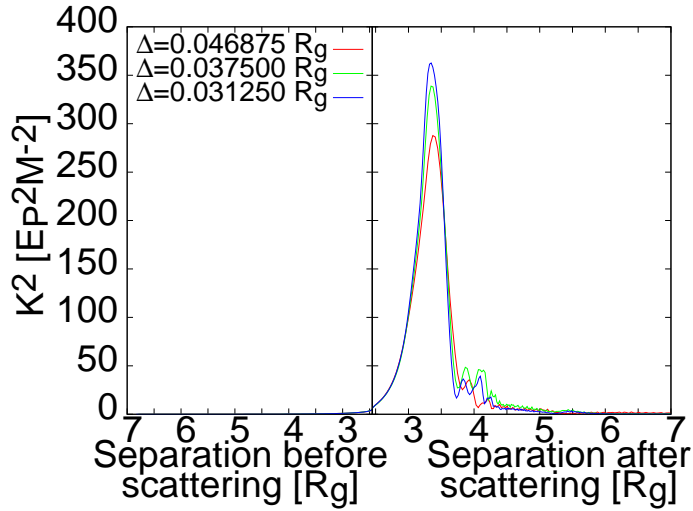}
 \label{fig:fig3}
 \caption{${\cal K}^2$ at the center of mass as a function of the
 coordinate separation between two black holes.  The parameters $v$
 and $b$ are the same as those for Fig.~2.  The time proceeds from
 left to right. The results with three grid resolutions are plotted.}
\end{figure}

Figure~3 plots ${\cal K}^2$ at the center of mass $w=x=y=z=0$ as a
function of the coordinate separation between two black holes. 
In this figure, the time elapses from left to right.  Initial speed of
each black hole and the impact parameter are the same as those for
Fig.~2.  As the separation between two black holes becomes small, the
value of $\cal K$ steeply increases. Then, after the black holes
slightly goes through the periastron, the maximum value of $\cal K$ is
reached. The maximum value ${\cal K}_{\rm max}$ in the best resolution
run is
\begin{equation}
{\cal K}_{\rm max}\simeq 19 \left(\frac{E_P}{M}\right).
\label{Kmax}
\end{equation}
In this scattering process, the event horizons of these black
holes do not merge with each other, because these black holes go away 
toward infinity separately after this scattering. Hence, by the symmetry
of this system, the center of mass is not enclosed by the event
horizon. This fact implies that even if each black hole is a
classical object before the scattering, a visible super-Planckian
domain can emerge in the vicinity of the center of mass in 5D general
relativity. 


\begin{figure}[t]
 \includegraphics[scale=0.7]{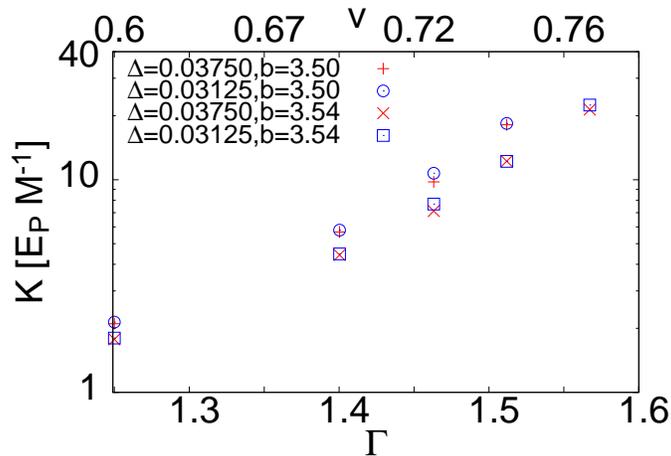} \label{fig:fig4} \caption{The
 maximum value of ${\cal K}$ calculated in the scattering of $b=3.5
 R_g$ and $b=3.54 R_g$ at the center of mass as a function of $v$.
 The results only for $b > b_{\rm C}$ are plotted.  The pluses and the
 crosses show the results of $\Delta/R_g=3/80$, and the circles and
 the squares show the results of $\Delta/R_g=1/32$.}
\end{figure}

Figure 4 plots the maximum values of $\cal K$ for the scattering of a
fixed impact parameter at the center of mass as a function of $v$. The
results only for $b > b_{\rm C}$ are plotted.  This shows that the
maximum value of ${\cal K}$ increases steeply with $v$, and thus, the
super-Planckian domain appears to be always visible in high-velocity
collisions. This also suggests that for $v \rightarrow 1$, the maximum
value of ${\cal K}$ would be much larger than $E_P/M$.

\section{Summary and Discussion}

Probing the short-distance physics would be limited if scattering
processes at energies well above the Planck scale were hidden behind a
black hole horizon.  If this hypothesis is correct, possible
productions of small black holes in particle colliders, such as the
CERN Large Hadron Collider, could be probably the unique opportunity
for exploring the nature of the quantum
gravity~\cite{GT01,DL01}. However, as shown in this paper, a visible
super-Planckian domain can be generated even through a scattering
process of two classical black holes in 5D spacetime.  We may say in
the practical sense that the cosmic censorship does not hold in 5D
spacetime, because a super-Planckian domain should be regarded as an
effective singularity in the classical gravity.

The visible super-Planckian domain is formed even for $v \sim 0.7$,
i.e., in a mildly relativistic scattering, and visible super-Planckian
domains can be generated through gravitational scatterings more
efficiently than a guess by NHM. It should be emphasized that such
domain could emerge even in the absence of special spacetime symmetry or
special assumption of spacetime geometry.  The present result implies
that it is necessary to study quantum gravitational effects in the
super-Planckian domain of no horizon.  Although, at present, we do not
know what happens in the super-Planckian domain, we expect that the
semi-classical particle creation in the sub-Planckian region could
occur around the visible super-Planckian domain~\cite{MNS2010}.

Finally, we comment on the scatterings for $b_{\rm B}(v)< b<b_{\rm
C}(v)$ with $v>0.6$ for which our numerical simulations do not keep
sufficient accuracy probably due to the emergence of a very large
spacetime curvature.  We performed simulations for $b > b_{\rm C}$
with several grid resolutions. The maximum value of $\cal K$ becomes
larger for the finer grid resolution for a given value of $b$. It also
increases with approaching $b_{\rm C}$.  These facts suggest that
ultra-high spacetime curvature, which is not encompassed in a horizon,
may be realized at some impact parameter for $b_{\rm B} < b < b_{\rm
C}$, i.e., the formation of a naked singularity.  We would like to
explore this issue in more detail elsewhere.

\vskip0.5cm
\noindent
{\bf Acknowledgements}

We thank T.~Tanaka for helpful duscussions.  KN is grateful to
C-M. Yoo, H.~Ishihara, and colleagues in astrophysics and gravity group
at Osaka City University for helpful discussions.  HO thanks
S. Kinoshita and N. Sago, and colleagues in astrophysics and cosmology
group at Yukawa Institute for Theoretical Physics for helpful
discussions.  This work was supported by Grant-in-Aid for Scientific
Research (21340051) and by Grant-in-Aid for Scientific Research on
Innovative Area (20105004) of Japanese MEXT.

\end{document}